\documentclass[twocolumn,showpacs,preprintnumbers,amsmath,amssymb]{revtex4}
\usepackage{amssymb}
\usepackage{graphicx}% Include figure files
\usepackage{dcolumn}% Align table columns on decimal point
\usepackage{bm}% bold math
\usepackage{textcomp}

\newcommand\ket[1]{\ensuremath{|#1\rangle}}

\newcommand\oprod[2]{\ensuremath{|#1\rangle\langle#2|}}

\begin{document}
\title{Tightened estimation can improve the key rate of MDI-QKD by more than 100\%}
\author{Yi-Heng Zhou$ ^{1}$,Zong-Wen Yu$ ^{2}$,
and Xiang-Bin Wang$ ^{1,3\footnote{Email
Address: xbwang@mail.tsinghua.edu.cn}}$}

\affiliation{ \centerline{$^{1}$State Key Laboratory of Low
Dimensional Quantum Physics, Tsinghua University, Beijing 100084,
People¡¯s Republic of China}\centerline{$^{2}$Data Communication Science and Technology Research Institute, Beijing 100191, China}\centerline{$^{3}$ Shandong
Academy of Information and Communication Technology, Jinan 250101,
People¡¯s Republic of China}}

%%%%%%%%%%%%%%%%%%%%%%%%%%%%%%%%%%%%%%%%%%%%%%%%%%%%%%%%%%%%%%%%%%%
%%%%%%%%%%%%%%%%%%%%%%%%%%%%%%%%%%%%%%%%%%%%%%%%%%%%%%%%%%%%%%%%%%%
%%%%%%%%%%%%%%%%%%%%%%%%% Abstract %%%%%%%%%%%%%%%%%%%%%%%%%%%%%%%%
\begin{abstract}
We present formulas to tightly upper bound the phase-flip errors in decoy state method by using 4 intensities. Our result compressed the bound to about a quarter of known result for MDI-QKD. Based on this, we find that the key rate is improved by more than 100\% given weak coherent state sources (WCS), and even more than 200\% with the heralded single-photon sources (HSPS).
\end{abstract}

%%%%%%%%%%%%%%%%%%%%%%%%%%%%%%%%%%%%%%%%%%%%%%%%%%%%%%%%%%%%%%%%%%%
%%%%%%%%%%%%%%%%%%%%%%%%%%%%%%%%%%%%%%%%%%%%%%%%%%%%%%%%%%%%%%%%%%%
%%%%%%%%%%%%%%%%%%%%%%%%%%%%%%%%%%%%%%%%%%%%%%%%%%%%%%%%%%%%%%%%%%%

\pacs{
03.67.Dd,
%Quantum cryptography
42.81.Gs,
%Birefringence, polarization
03.67.Hk
%Quantum communication
}
\maketitle

%%%%%%%%%%%%%%%%%%%%%%%%%%%%%%%%%%%%%%%%%%%%%%%%%%%%%%%%%%%%%%%%%%%
%%%%%%%%%%%%%%%%%%%%%%%%%%%%%%%%%%%%%%%%%%%%%%%%%%%%%%%%%%%%%%%%%%%
%%%%%%%%%%%%%%%%%%%%%%%%%%%%%%%%%%%%%%%%%%%%%%%%%%%%%%%%%%%%%%%%%%%
%%%%%%%% Introducation & Motivation %%%%%%%%%%%%%%%%%%%%%%%%%%%%%%%

\section{Introduction}\label{SecIntro}
One of the most fascinating properties of quantum key distribution (QKD) is its unconditional security in theory~\cite{BB84,GRTZ02}. However, most practical devices behave differently form the theoretical models assumed in the security proof. Security for real set-ups of QKD~\cite{BB84,GRTZ02} has become a major problem in this area in the recent years. The insecurity loopholes are mainly due to the imperfect single-photon source and the limited efficiency of the detectors. Fortunately, by using the decoy-state method~\cite{ILM,H03,wang05,LMC05,AYKI,haya,peng,wangyang,rep,njp}, it has been shown that the unconditional security of QKD can still be assured with an imperfect single-photon source~\cite{PNS1,PNS}.

Besides the source imperfection, the defect in the detectors is another threaten to the security~\cite{lyderson}.
To patch up this, several approaches have been proposed. One is the device independent QKD (DI-QKD)~\cite{ind1}. This technique does not require detailed knowledge of how QKD devices work and can prove security based on the violation of a Bell inequality.

Recently, an idea of measurement device independent QKD (MDI-QKD) was proposed based on the idea of entanglement swapping~\cite{ind3,ind2}. There, one can make secure QKD simply by virtual entanglement swapping, i.e., neither Alice and Bob performs any measurement, but they only send out quantum signals to the relay which can be controlled by the un-trusted third party (UTP). After Alice and Bob send out signals, they wait for UTP's announcement of weather he has obtained the successful detection, and proceed to the standard postprocessing of their sifted data, such as error rate estimation, error correction, and privacy amplification. The only assumption needed in MDI-QKD is that the preparation of the quantum signal sources by Alice and Bob. In practice, in order to obtain a higher key rate or realize a longer distance key distribution, we'd better use laser sources with decoy state method. This has been discussed in Ref.~\cite{ind2}, and explicit formulas for the practical decoy-state implementation with only three different states was first presented in \cite{wangPRA2013}, and then further studied  both experimentally~\cite{tittel1,tittel2,liuyang} and theoretically\cite{wangArxiv,lopa,curtty,Wang3int,Wang3improve,Wang3g,WangModel}. In the previous works, the authors considered the effect of finite number of decoy states, but their key rates are notably away form the result obtained with the infinite decoy-state method. The major reason is that the upper bounds of the phase-flip error estimated with these methods are not very tightened.

Here in this work, we show how to tightly formulate the upper bound of the phase-flip errors in decoy state method for the regular BB84 protocol and MDI-QKD. Our result compressed the bound to about a quarter of known result for MDI-QKD with WCS, and even about one fifth with HSPS. To achieve the result, we only need 4-intensity decoy state method. Based on this, we find that the key rate is improved by more than 100\% with WCS, and even more than 200\% with HSPS.

\section{Traditional Decoy-state method with only 4 intensities for BB84 protocol}\label{SecBB84}
In the four-intensity protocol, Alice has four (virtual) sources, the vacuum source $\rho_0=\oprod{0}{0}$ which prepares vacuum pulses, two decoy sources $\rho_{x},\rho_{y}$ which prepare decoy pulses, and the signal source $\rho_{z}$ which prepares signal pulses. In photon-number space, we suppose
\begin{equation}\label{BB84Rhoxyz}
  \rho_{l}=\sum_{k}a_k^l \oprod{k}{k}, \quad (l=x,y,z),
\end{equation}
where $\ket{k}$ is the $k$-photon Fock state, $a_k^{l}\geq 0$ for all $k\geq 0$.

At each time, Alice will randomly select one of her 4 sources to emit a pulse. For pose-processing, Alice and Bob evaluate the data with the same basis. With the observed total gains and error rates, the final secure key rate can be calculated by the following formula~\cite{ILM}
\begin{equation}\label{BB84R}
  R=a_1^z s_1[1-H(e_1)]-S_z H(E_z),
\end{equation}
where $S_z$ and $E_z$ denote, respectively, the total gain and error rate of the signal state $\rho_z$. $s_1$ and $e_1$ are, respectively, the fraction and error rate of detection events by Bob that have originated form single-photon pulses emitted by Alice, and $H$ is the binary Shannon entropy. In this paper, we use the capital letter $S(E)$ for {\em{known}} total gains (error rates) and the lowercase letter $s,e$ for unknown variables.

In order to estimate the final key rate of this protocol, we need find out the lower bound of the yield $s_1$ and the upper bound of the error rate $e_1$. In the coming subsection, we devote to estimate the lower bound of $s_1$ firstly.

\subsection{The lower bound of the yield $s_1$}
With given the different sources, Alice randomly chooses quantum channels with different photon-number states. Thus, the total gain with source $\rho_l$ can be expressed into the following convex form
\begin{equation}\label{BB84Rhol}
  S_{l}=\sum_{k\geq 0}a_k^{l}s_k, \quad (l=x,y,z),
\end{equation}
where $s_k$ is the yield of an $k$-photon pulse. In order to obtain an effective lower bound of $s_1$, we need eliminate the gains associated with the vacuum state from the total gain firstly. Considering this fact, we can rewrite the relations in Eq.(\ref{BB84Rhol}) into
\begin{equation}\label{BB84Rhot}
  \tilde{S}_l=\sum_{k\geq 1}a_k^{l}s_k, \quad (l=x,y,z),
\end{equation}
where we define
\begin{equation}\label{BB84TS}
  \tilde{S}_l=S_l-a_0^l S_0, \quad (l=x,y,z),
\end{equation}
with $S_0$ being the gain of the vacuum source.

The lower bound of $s_1$ has already been studied~\cite{wang05,LMC05,AYKI,haya}. As presented in the previous works, if Alice has 3 different sources $\rho_{o}=\oprod{0}{0},\rho_x,\rho_y$, the lower bound of $s_1$ can be write into
\begin{equation}\label{BB84s1L3}
  \underline{s}_1(x,y)=\frac{a_2^y \tilde{S}_x -a_2^x\tilde{S}_y}{a_1^x a_2^y -a_1^y a_2^x},
\end{equation}
under the condition
\begin{equation}\label{BB84Condxy}
  \frac{a_k^y}{a_k^x}\geq \frac{a_2^y}{a_2^x}\geq \frac{a_1^y}{a_1^x},
\end{equation}
for all $k\geq 2$. It is worth pointing out that the lower bound given by Eq.(\ref{BB84s1L3}) does not only apply to the weak coherent source, but also to any source as long as it meets the condition in Eq.(\ref{BB84Condxy}).

In this 4-intensity protocol, there are three different no-vacuum sources. In order to get the lower bound of $s_1$ directly from Eq.(\ref{BB84s1L3}), we also need to introduce the following condition
\begin{equation}\label{BB84Condyz}
  \frac{a_k^z}{a_k^y}\geq \frac{a_2^z}{a_2^y}\geq \frac{a_1^z}{a_1^y},
\end{equation}
for all $k\geq 2$. Then we can obtain some effective lower bounds of $s_1$ with Eq.(\ref{BB84s1L3}) by choosing any two different sources from $\rho_x,\rho_y$ and $\rho_z$. After this, we can use the maximum one as the estimation of the lower bound of $s_1$ for this 4-intensity protocol
\begin{equation}\label{BB84s1L4}
  \underline{s}_1^{\prime}=\max\{\underline{s}_1(x,y),\underline{s}_1(x,z),\underline{s}_1(y,z)\},
\end{equation}
where $\underline{s}_1(l,r)$ is just $\underline{s}_1(x,y)$ in Eq.(\ref{BB84s1L3}) with changing $x$ and $y$ into $l$ and $r$ respectively. In order to simplify this expression and derive other main results in this work, we need to define the following function with sources $\rho_x,\rho_y$ and $\rho_z$
\begin{equation}\label{Gdef}
  \mathcal{G}(i,j,k)=(g_i^x-g_j^x)(g_j^y-g_k^y)-(g_i^y-g_j^y)(g_j^x-g_k^x),
\end{equation}
where
\begin{equation}\label{gdef}
  g_m^l=\frac{a_m^l}{a_m^z}, \quad (m\geq 1;\,l=x,y,z).
\end{equation}
Now, we assume that the states $\rho_x,\rho_y$ and $\rho_z$ satisfy the following important condition
\begin{equation}\label{BB84Condxyz}
  \mathcal{G}(i,j,k)\geq 0,
\end{equation}
when $k-j\geq j-i\geq 0$. In Appendix A, we will show that the imperfect sources used in practice such as the weak coherent sources, the heralded source out of the parametric-down conversion, satisfy all the above conditions given by Eqs.(\ref{BB84Condxy},\ref{BB84Condyz},\ref{BB84Condxyz}). With these conditions presented in Eqs.(\ref{BB84Condxy},\ref{BB84Condyz},\ref{BB84Condxyz}), we can simplify the lower bound $\underline{s}_1^{\prime}$ by
\begin{equation}\label{BB84s1L4s}
  \underline{s}_1^{\prime}=\underline{s}_1(x,y).
\end{equation}
The detailed proof of this conclusion can be found in Appendix B.
%In the third subsection of this part, we will show some numerical simulations. In Fig.\ref{BB84s1L}, we can find out that the lower bound of $s_1$ is very close to the results obtained with the infinite-decoy method.

\subsection{The upper bound of the error rate $e_1$}
In order to estimate the final key rate, we also need the upper bound of the error rate $e_1$. In the previous works, the upper bound of $e_1$ is obtained by putting the errors with all muti-photon pulses on the error with the single-photon pulse. Explicitly, we can write the upper bound of $e_1$ with 3-intensity decoy state method as follows
\begin{equation}\label{BB84e1U3}
  \overline{e}_1=\frac{S_x E_x-a_0^x S_0 E_0}{a_1^x \underline{s}_1},
\end{equation}
where $\underline{s}_1$ is the lower bound of $s_1$, $S_x$ and $E_x$ are the total gain and error rate of the source $\rho_x$ respectively, $S_0$ and $E_0$ are the total gain and error rate of the vacuum source respectively. With the 3-intensity decoy state method, we can not find out a more better explicit formula to estimate the upper bound of $e_1$. In order to get a more tightened upper bound, we need to introduce one more source. This is the main reason for us to consider the 4-intensity decoy state method.

Similar to the gain, the error rate can depend on the photon number. Let us denote $e_k$ as the error of an $k$-photon pulse. The error rate $E_l$ for the source $\rho_{l}(l=x,y,z)$ can be given by
\begin{equation}\label{BB84E4}
  T_l=S_l E_l=\sum_{k\geq 0} a_k^{l} s_k e_k, \quad (l=x,y,z).
\end{equation}
If we denote $t_k=s_k e_k$, and
\begin{equation}\label{BB84TE4def}
  \tilde{T}_{l}=T_l-a_0^{l} T_0, \quad (l=x,y,z),
\end{equation}
Eq.(\ref{BB84E4}) can be rewrite into the following equivalent form
\begin{equation}\label{BB84TE4}
  \tilde{T}_{l}=\sum_{k\geq 1} a_k^{l} t_k, \quad (l=x,y,z).
\end{equation}

In this 4-intensity decoy state method, there are 3 different no-vacuum sources can be used for Alice. Then we have 3 different relations about $t_1$ which are presented in Eq.(\ref{BB84TE4}). With these 3 relations, by eliminating the variables $t_2$ and $t_3$, we obtain the expression of $t_1$ as follows
\begin{equation}\label{BB84t1}
  t_1=\overline{t}_1^{\prime}+\sum_{k\geq 4}f_{t_1}(k)t_k,
\end{equation}
where
\begin{widetext}
\begin{equation}\label{BB84t1U}
  \overline{t}_1^{\prime}=\frac{a_1^z a_2^z a_3^z}{\mathcal{G}(1,2,3)}\left[(a_3^z a_2^y- a_3^y a_2^z)\tilde{T}_x-(a_3^z a_2^x-a_3^x a_2^z)\tilde{T}_y +(a_3^y a_2^x -a_3^x a_2^y)\tilde{T}_z\right],
\end{equation}
\end{widetext}
and
\begin{equation}\label{BB84ft1k}
  f_{t_1}(k)=-\frac{\mathcal{G}(2,3,k)}{\mathcal{G}(1,2,3)}, \quad (k\geq 4),
\end{equation}
with $\mathcal{G}(1,2,3)$ being defined in Eq.(\ref{Gdef}). Under the condition presented in Eq.(\ref{BB84Condxyz}), we can easily find out that $f_{t_1}(k)\leq 0$ for all $k\geq 4$. Then we can conclude that $\overline{t}_1^{\prime}$ given by Eq.(\ref{BB84t1U}) is actually a upper bound of $t_1$. Then the upper bound of $e_1$ can be given by
\begin{equation}\label{BB84e1U4}
  \overline{e}_1^{\prime}=\frac{\overline{t}_1^{\prime}}{\underline{s}_1^{\prime}}
\end{equation}
where $\underline{s}_1^{\prime}$ is the lower bound of $s_1$ given by Eq.(\ref{BB84s1L4s}).

\subsection{Numerical Simulation for BB84 protocol}\label{SubSection2C}

\begin{figure}
  \includegraphics[width=240pt]{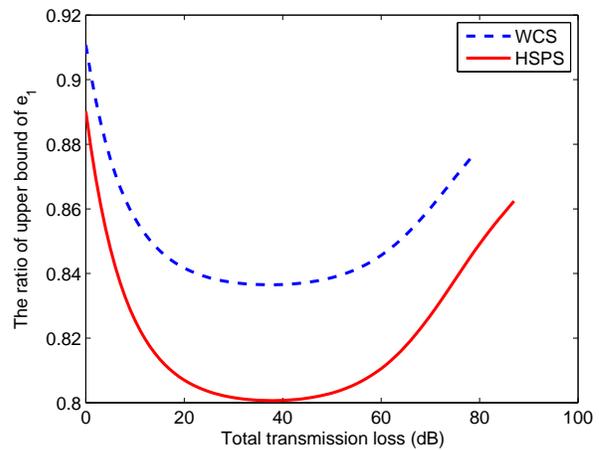}\\
  \caption{(Color online) The ratio of the upper bound of $e_{1}$ between the estimations obtained by using 4-intensity and 3-intensity decoy state methods, i.e., $\overline{e}_1^{\prime}/\overline{e}_1$, versus the total channel transmission loss. We set $\mu_1=0.2$ for decoy state.}\label{BB84e1U4over3}
\end{figure}

\begin{figure}
  \includegraphics[width=240pt]{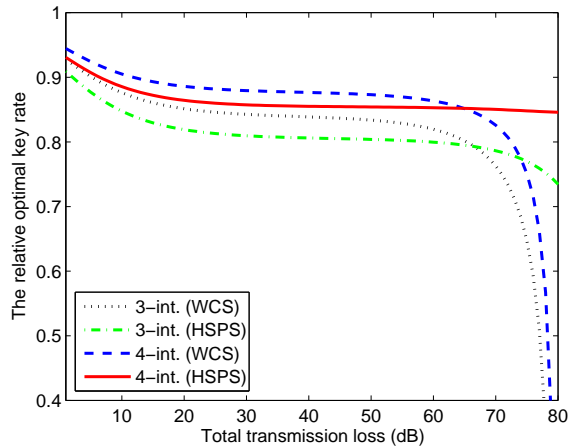}\\
  \caption{(Color online) The relative value between the optimal key rate obtained with different methods and the asymptotic limit of the infinite decoy-state method versus the total channel transmission loss. We set $\mu_1=0.2$ for  decoy states.}\label{BB84rROpt}
\end{figure}

\begin{figure}
  \includegraphics[width=240pt]{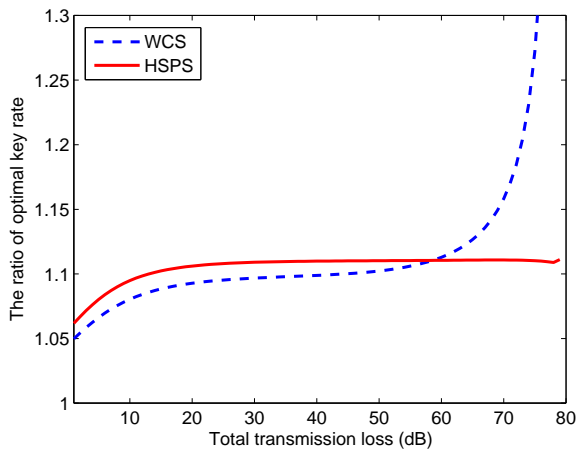}\\
  \caption{(Color online) The ratio of the optimal key rate between the estimations obtained by using 4-intensity and 3-intensity decoy state methods versus the total channel transmission loss. We set $\mu_1=0.2$ for  decoy state.}\label{BB84ROpt4over3}
\end{figure}

\begin{table}%[!hbp]
\caption{\label{BB84table}List of experimental parameters used in numerical simulations: $e_0$ is the error rate of background, $e_d$ is the misalignment-error probability; $p_d$ is the dark count rate of Bob's per detector; $\eta_v$ is the detection efficiency of Alice's detector; $p_{dv}$ is the dark count rate of Alice's detector.}
\begin{ruledtabular}
\begin{tabular}{ccccc}
  $e_0$ & $e_d$ & $p_d$ & $\eta_v$ & $p_{dv}$  \\
  \hline
  0.5 & 1.5\% & $3.0\times 10^{-6}$ & 0.75 & $1.0\times 10^{-6}$ \\
\end{tabular}
\end{ruledtabular}
\end{table}

In this subsection, we will present some numerical simulations to compare the results obtained by using the 3-intensity decoy state method with the results of 4-intensity method for the regular BB84 protocol. As discussed before, we know that the methods presented in this work does not only apply to the weak coherent sources (WCS). Actually, it can be used to estimate the final key rate for any sources that satisfy the condition given by Eq.(\ref{BB84Condxy}) for the 3-intensity method, and the conditions given by Eqs.(\ref{BB84Condxy},\ref{BB84Condyz},\ref{BB84Condxyz}) for the 4-intensity method. Below for simplicity, we consider the following two cases. In the first case, we suppose that Alice use WCS. In the second one, we suppose she use the heralded single-photon sources (HSPS) with possion distributions~\cite{wangArxiv}. The Bob's detectors are identical, i.e., they have the same dark count rate and detection efficiency, and the detection efficiency does not depend on the incoming states. Suppose the overall transmission probability of each photon is $\xi$. In a normal channel, it is common to assume independence between the behaviors of the $n$ photons. Therefore, the transmission efficiency for $n$-photon pulses $\xi_n$ is given by
\begin{equation*}
  \xi_n=1-(1-\xi)^n.
\end{equation*}
For fair comparison, we use the same parameter values used in~\cite{UrsinNP2007} for our numerical evaluation. For simplicity, we shall put the detection efficiency to the overall transmittance $\eta=\xi\zeta$. We assume all detectors of Bob have the same detection efficiency $\zeta$ and dark count rate $p_d$. In the second case with HSPS, we assume the detector of Alice has the detection efficiency $\eta_v$ and dark count rate $p_{dv}$. The values of these parameters are presented in Table~\ref{BB84table}. With this, the total gains $S_{\mu_i}$ and error rates $S_{\mu_i}E_{\mu_i}$ of Alice's intensity $\mu_i$ ($i=0,1,2$ for 3-intensity method, $i=0,1,2,3$ for 4-intensity method) can be calculated. By using these values, we can estimate the lower bounds of yield $s_1$ with Eq.(\ref{BB84s1L3}) and Eq.(\ref{BB84s1L4s}) for 3-intensity and 4-intensity decoy state methods respectively. Also, we can estimate the upper bounds of error rate $e_1$ with Eq.(\ref{BB84e1U3}) and Eq.(\ref{BB84e1U4}) for 3-intensity and 4-intensity decoy sate methods respectively. Furthermore, with these parameters, we can estimate the final key rate $R$ of this protocol with Eq.(\ref{BB84R}). If we fix the density(ies) of the decoy-state source(s) used by Alice, the final key rate will change with Alice taking different intensities for hers signal-state pulses. Here, in order to make a rational and effective comparison, we set the intensities of the decoy source in 3-intensity method and the first decoy source in 4-intensity method are the same and $\mu_1=0.2$; let the intensity of the second decoy source in 4-intensity method to be the optimal intensity of signal source in 3-intensity method and assume $\mu_2>\mu_1$.

With these preparations, we can conclude that the lower bounds of $s_1$ estimated by using the 3-intensity and 4-intensity decoy state methods are the same, i.e., $\underline{s}_1=\underline{s}_1^{\prime}$. In order to see more clearly, in Fig.\ref{BB84e1U4over3}, we plot the ratio of the upper bound of $e_{1}$ between the estimations obtained by using 4-intensity and 3-intensity decoy state methods, i.e., $\overline{e}_1^{\prime}/\overline{e}_1$. The relative value between the optimal key rate obtained with different methods and the asymptotic limit of the infinite decoy-state method are shown in Fig.\ref{BB84rROpt}. In order to clarify the superiority of the 4-intensity decoy state method, we plot the ratio of the optimal key rate between the results obtained by using 4-intensity and 3-intensity decoy state methods in Fig.\ref{BB84ROpt4over3}. In Fig.\ref{BB84e1U4over3} and Fig.\ref{BB84ROpt4over3}, the blue dashed lines are obtained with WCS, the red solid lines are obtained with HSPS. In Fig.\ref{BB84rROpt}, the black dotted line and green dash-dot line are the results obtained by using 3-intensity decoy state method with WCS and HSPS respectively, the blue dashed line and the red solid line are the results obtained by using 4-intensity decoy state with WCS and HSPS respectively. With these 3 figures, we can conclude that the results obtained by using the 4-intensity decoy state method are better than the results of 3-intensity method. But it only have a litter improvement.

\section{Tightened formula for decoy-state MDI-QKD with 4 intensities}\label{SecMDI}
In the protocol, each time a pulse-pair (two-pulse state) is sent to the relay for detection. The relay is controlled by an UTP. The UTP will announce whether the pulse-pair has caused a successful event. Those bits corresponding to successful events will be post-selected and further processed for the final key. Since real set-ups only use imperfect single-photon sources, we need the decoy-state method for security.

We assume Alice (Bob) has four sources, $o_A,x_A,y_A,z_A$ ($o_B,x_B,y_B,z_B$) which can only emit four different states $\rho_{r_A}(\rho_{r_B})$,$(r=o,x,y,z)$. In the following discussion, we assume $o_A$ and $o_B$ are two vacuum sources. In photon number space, we have $\rho_{o_A}=\oprod{0}{0}, \rho_{o_B}=\oprod{0}{0}$. For the others, suppose
\begin{equation*}\label{MDIsources}
  \rho_{r_A}=\sum_{k} a_{k}^{r}\oprod{k}{k}, \rho_{r_B}=\sum_{k} b_k^{r}\oprod{k}{k},(r=x,y,z).
\end{equation*}
In order to obtain the main results, we also need to introduce the following function
\begin{equation}\label{Hdef}
  \mathcal{H}(i,j,k)=(h_i^x-h_j^x)(h_j^y-h_k^y)-(h_i^y-h_j^y)(h_j^x-h_k^x),
\end{equation}
where
\begin{equation}\label{hdef}
  h_n^l=\frac{b_n^l}{b_n^z}, \quad (n\geq 1,\,l=x,y,z).
\end{equation}
Now, we assume that the states $\rho_{x_{A(B)}},\rho_{y_{A(B)}}$ and $\rho_{z_{A(B)}}$ satisfy the following important conditions:
\begin{equation}\label{MDIcondxy}
  \frac{c_k^z}{c_k^y} \geq \frac{c_2^z}{c_2^y} \geq \frac{c_1^z}{c_1^y}, \quad
  \frac{c_k^y}{c_k^x} \geq \frac{c_2^y}{c_2^x} \geq \frac{c_1^y}{c_1^x}, \quad (c=a,b),
\end{equation}
for $k\geq 2$, and
\begin{equation}\label{MDIcondxyz}
  \mathcal{G}(i,j,k)\geq 0, \quad \mathcal{H}(i,j,k)\geq 0,
\end{equation}
when $k-j\geq j-i\geq 0$. Similar to $\mathcal{G}$, the imperfect sources used in practice such as the coherent state source, the heralded source out of the parametric-down conversion, satisfy the above restrictions. Given a specific type of source, the above listed different states have different averaged photon numbers (intensities), therefore the states can be obtained by controlling the light intensities.

At each time, Alice will randomly select one of her 3 sources to emit a pulse, and so does Bob. The pulse from Alice and the pulse from Bob form a pulse pair and are sent to the un-trusted relay.  We regard equivalently that each time a two-pulse source is selected and a pulse pair (one pulse from Alice, one pulse from Bob) is emitted. For post-processing, Alice and Bob evaluate the data sent in two bases separately. The $Z$-basis is used for key generation, while the $X$-basis is used for testing against tampering and the purpose of quantifying the amount of privacy amplification needed. With the observed total gains and error rates, we can calculate the final secure key rate with the following formula~\cite{ind2}
\begin{equation}\label{KeyRate}
  R=a_1^z b_1^z s_{11}^{Z}[1-H(e_{11}^{X})]-S_{z_A z_B}^{Z}fH(E_{z_A z_B}^{Z}),
\end{equation}
where $S_{z_A z_B}^{Z}$ and $E_{z_A z_B}^{Z}$ denote, respectively, the gain and error rate in the $Z$-basis when both Alice and Bob use $z$-source $\rho_{z_A}$ and $\rho_{z_B}$; $f$ is the efficiency factor of the error correction method used; $s_{11}^{Z}$ and $e_{11}^{X}$ are the gain and error rate when both Alice and Bob send single-photon states. In this paper, we use capital letter $Z(X)$ for the bases and the lowercase letter $x,y,z$ for the different sources.

In order to estimate the final key rate of this protocol, we need find out the lower bound of the yield $s_{11}$ and the upper bound of the error rate $e_{11}$.

\subsection{The lower bound of the yield $s_{11}$}
With given the different sources, Alice and Bob randomly choose quantum channels with different photon-number states. Thus, the total gain can be expressed into the following convex form
\begin{equation}\label{GainAB}
   S_{lr}=\sum_{j,k\geq 0}a_{j}^{l}b_{k}^{r}s_{jk}, \quad (l,r=x,y,z),
\end{equation}
when Alice and Bob send pulses with $\rho_{l_A}$ and $\rho_{r_B}$ respectively. Here and after, we omit the subscripts $A$ and $B$ without causing any ambiguity. It is well-known that, in order to obtain an effective lower bound of $s_{11}$, we need eliminate the gains associated with the vacuum state from the total gain firstly. Considering this fact, we can rewrite the relation in Eq.(\ref{GainAB}) into
\begin{equation}\label{tGainAB}
  \tilde{S}_{lr}=\sum_{j,k\geq 1}a_{j}^{l}b_{k}^{r}s_{jk}, \quad (l,r=x,y,z),
\end{equation}
with
\begin{equation}\label{tSdef}
  \tilde{S}_{lr}=S_{lr}-a_0^{l}S_{0r}-b_0^{r}S_{l0}+a_0^{l}b_0^{r}S_{00}, \quad (l,r=x,y,z).
\end{equation}

The lower bound of $s_{11}$ has already been exhaustive studied for 3-intensity decoy state MDI-QKD protocol~\cite{wangPRA2013,wangArxiv,Wang3int,Wang3improve,Wang3g}. Until now, the most tightly explicit formula to calculate the lower bound of $s_{11}$ is given in Ref.\cite{Wang3int}. As presented in Ref.\cite{Wang3int}, the lower bound of $s_{11}$ with 3 different sources ($o_A,l_A,r_A$ and $o_B,l_B,r_B$) used in each side of Alice and Bob can be expressed as
\begin{widetext}
\begin{equation}\label{s11L3int}
  \underline{s}_{11}(l,r)=\frac{(a_1^l a_2^r b_1^l b_2^r-a_1^r a_2^l b_1^r b_2^l)\tilde{S}_{ll}-b_1^l b_2^l(a_1^l a_2^r-a_1^r a_2^l)\tilde{S}_{lr}-a_1^l a_2^l (b_1^l b_2^r-b_1^r b_2^l)\tilde{S}_{rl}}{a_1^l b_1^l(a_1^l a_2^r-a_1^r a_2^l)(b_1^l b_2^r-b_1^r b_2^l)},
\end{equation}
\end{widetext}
under the condition
\begin{equation*}
  \frac{a_k^r}{a_k^l}\geq \frac{a_2^r}{a_2^l}\geq \frac{a_1^r}{a_1^l}, \quad
  \frac{b_k^r}{b_k^l}\geq \frac{b_2^r}{b_2^l}\geq \frac{b_1^r}{b_1^l},
\end{equation*}
for all $k\geq 2$, where $\tilde{S}_{ll},\tilde{S}_{lr},\tilde{S}_{rl}$ are the amended gains defined by Eq.(\ref{tGainAB}). In this 4-intensity protocol, there are 3 no-vacuum sources. We can estimate the effective lower bounds of $s_{11}$ with Eq.(\ref{s11L3int}) by choosing $l$ and $r$ as any two different sources from $x,y,z$. Then we can use the maximum one as the lower bound of $s_{11}$ for this 4-intensity protocol
\begin{equation}\label{s11L4int}
  \underline{s}_{11}^{\prime}=\max\{\underline{s}_{11}(x,y),\underline{s}_{11}(x,z),\underline{s}_{11}(y,z)\}.
\end{equation}
Actually, under the assumptions given by Eqs.(\ref{MDIcondxy}-\ref{MDIcondxyz}), we can simplify the lower bound of $s_{11}$ in Eq.(\ref{s11L4int}) by choosing the lowest two sources at each sides of Alice and Bob, such that
\begin{equation}\label{s11L4ints}
  \underline{s}_{11}^{\prime}=\underline{s}_{11}(x,y).
\end{equation}
The detailed proof of this conclusion can be found in appendix B.

\subsection{The upper bound of the error rate $e_{11}$}
In order to estimate the final key rate, we also need the upper bound of the error rate $e_{11}$. In previous works, the upper bound of $e_{11}$ is obtained by putting the errors with all multi-photon pairs on the error with the single-photon pair. Explicitly, we can write the upper bound of $e_{11}$ with 3-intensity decoy state method as follows
\begin{equation}\label{e11U3int}
  \overline{e}_{11}=\frac{S_{xx} E_{xx}-a_0^x S_{0x} E_{0x}-b_0^x S_{x0}E_{x0}+a_0^x b_0^x S_{00}E_{00}}{a_1^x b_1^x \underline{s}_{11}},
\end{equation}
where $\underline{s}_{11}$ is the lower bound of $s_{11}$, $S_{lr}$ and $E_{lr}$ are the total gain and error rate when Alice use the source $\rho_{l_A}$ and Bob use the source $\rho_{r_B}$ respectively. With the numerical results presented in the third subsection of this part, we know that the upper bound obtained with this method is too rough to get an tight estimation of the final key rate comparing with the results obtained by using the infinite-decoy sate method. In order to find out a more tightened upper bound of $e_{11}$, we need introduce one more source in each side of Alice and Bob. This is the main reason for us to consider the 4-intensity decoy state method for MDI-QKD. As expected, we can find out a more tightened upper bound of $e_{11}$ for this protocol.

Similar to the total gain, the error rate can be write into the following convex expressions
\begin{equation}\label{tErrorAB}
  \tilde{T}_{lr}=\sum_{j,k\geq 1}a_{j}^{l}b_{k}^{r}t_{jk}, \quad (l,r=x,y,z),
\end{equation}
where $T_{lr}=E_{lr}S_{lr}$, $t_{jk}=s_{jk}e_{jk}$, and
\begin{equation}\label{tEdef}
  \tilde{T}_{lr}=T_{lr}-a_0^{l}T_{0r}-b_0^{r}T_{l0}+a_0^{l}b_0^{r}T_{00}, \quad (l,r=x,y,z).
\end{equation}
In this 4-intensity protocol, there are 3 different no-vacuum sources in each side of Alice and Bob. Then we have 9 different relations about $e_{11}$ which are given by Eq.(\ref{tErrorAB}). With these 9 relations, by eliminating the variables $t_{12},t_{21},t_{22},t_{13},t_{23},t_{33},t_{32},t_{31}$, we obtain the expression of $t_{11}$
\begin{equation}\label{t11Exp}
  t_{11}=\overline{t}_{11}^{\prime}+\sum_{(m,n)\in J_{0}}f_{t_{11}}(m,n)t_{mn},
\end{equation}
where $J_0=\{(m,n)|m,n\geq 1; m+n\geq 5;(m,n)\neq (2,3);(m,n)\neq (3,3);(m,n)\neq (3,2)\}$,
\begin{widetext}
\begin{equation}\label{t11U}
  \overline{t}_{11}^{\prime}=\frac{(a_2^y a_3^z- a_2^z a_3^y)\mathcal{T}_b^x-(a_2^x a_3^z- a_2^z a_3^x)\mathcal{T}_b^y+ (a_2^x a_3^y- a_2^y a_3^x)\mathcal{T}_b^z}{a_1^z a_2^z a_3^z \mathcal{G}(1,2,3) b_1^z b_2^z b_3^z \mathcal{H}(1,2,3)},
\end{equation}
%\begin{equation}\label{fmn}
%  f(m,n)=-\frac{a_m^z(h_3^y h_m^x -h_2^y h_m^x-h_3^x h_m^y +h_2^x h_m^y +h_2^y h_3^x-h_2^x h_3^y) b_n^z(g_3^y g_k^x %-g_2^y g_k^x-g_3^x g_k^y +g_2^x g_k^y +g_2^y g_3^x-g_2^x g_3^y)}{a_1^z H(3,2,1) b_1^z G(3,2,1)},
%\end{equation}
\end{widetext}
and
\begin{equation}\label{fmn}
  f_{t_{11}}(m,n)=-\frac{a_m^z\mathcal{G}^{\prime}(m) b_n^z\mathcal{H}^{\prime}(n)}{a_1^z \mathcal{G}(1,2,3) b_1^z\mathcal{H}(1,2,3)},
\end{equation}
with
\begin{equation*}
  \mathcal{T}_b^l=(h_2^y-h_3^y)\tilde{T}_{lx}-(h_2^x-h_3^x)\tilde{T}_{ly}+(h_3^y h_2^x-h_3^x h_2^y)\tilde{T}_{lz},
\end{equation*}
for $l=x,y,z$, and
\begin{eqnarray*}
  \mathcal{G}^{\prime}(m)&=&\left\{\begin{array}{cc}
    \mathcal{G}(m,2,3), & m=1,2; \\
    \mathcal{G}(2,3,m), & m\geq 3,
  \end{array}\right. \label{Ginfmn} \\
  \mathcal{H}^{\prime}(n)&=&\left\{\begin{array}{cc}
    \mathcal{H}(n,2,3), & n=1,2; \\
    \mathcal{H}(2,3,n), & n\geq 3.
  \end{array}\right. \label{Hinfmn}
\end{eqnarray*}
Here, $h_k^l$ is defined by Eq.(\ref{hdef}), $\mathcal{G}(i,j,k)$ and $\mathcal{H}(i,j,k)$ are defined in Eq.(\ref{Gdef}) and Eq.(\ref{Hdef}) respectively. With the conditions presented in Eqs.(\ref{MDIcondxy}-\ref{MDIcondxyz}), we can prove that
\begin{equation}\label{Propfmn}
  f_{t_{11}}(m,n)\leq 0,
\end{equation}
for all $(m,n)\in J_0$. Then we conclude that the expression given by Eq.(\ref{t11U}) is actually an upper bound of $t_{11}$. With this, we can estimate the upper bound of $e_{11}$ by the following explicit formula
\begin{equation}\label{e11U4int}
  \overline{e}_{11}^{\prime}=\frac{\overline{t}_{11}^{\prime}}{\underline{s}_{11}^{\prime}},
\end{equation}
where $\underline{s}_{11}^{\prime}$ is the lower bound of $s_{11}$ given in Eq.(\ref{s11L4ints}).

\subsection{Numerical Simulation for MDI-QKD}

\begin{figure}
  \includegraphics[width=240pt]{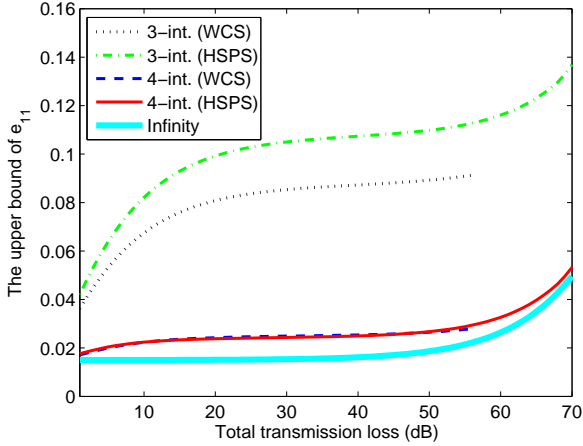}\\
  \caption{(Color online) The estimated values of $e_{11}$ versus the total channel transmission loss for MDI-QKD with WCS and HSPS. We set $\mu_1=\nu_1=0.1$ for decoy state, and $\mu_2=\nu_2$.}\label{MDIre11U}
\end{figure}

\begin{figure}
  \includegraphics[width=240pt]{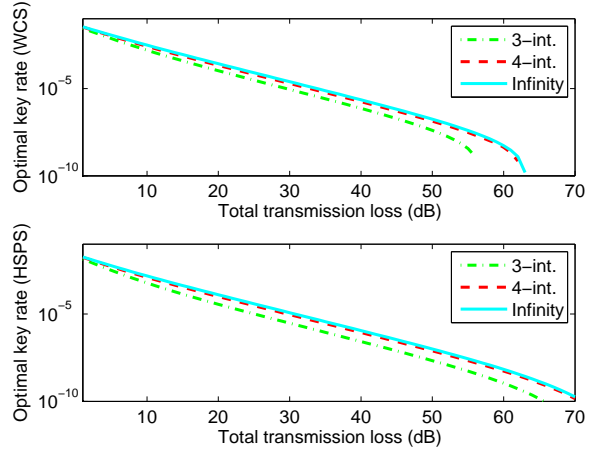}\\
  \caption{(Color online) The optimal key rate versus the total channel transmission loss using different methods for MDI-QKD with WCS and HSPS. We set $\mu_1=\nu_1=0.1$ for decoy states, and $\mu_2=\nu_2$.}\label{MDIROpt}
\end{figure}

\begin{figure}
  \includegraphics[width=240pt]{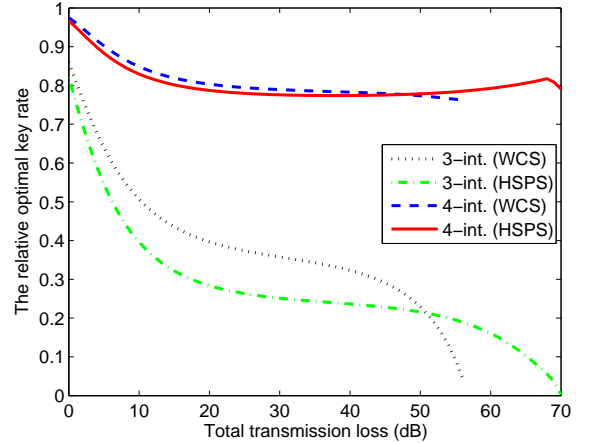}\\
  \caption{(Color online) The relative optimal key rate of different methods versus the total channel transmission loss for MDI-QKD with WCS and HSPS. We set $\mu_1=\nu_1=0.1$ for decoy states, and $\mu_2=\nu_2$.}\label{MDIrROpt}
\end{figure}

\begin{figure}
  \includegraphics[width=240pt]{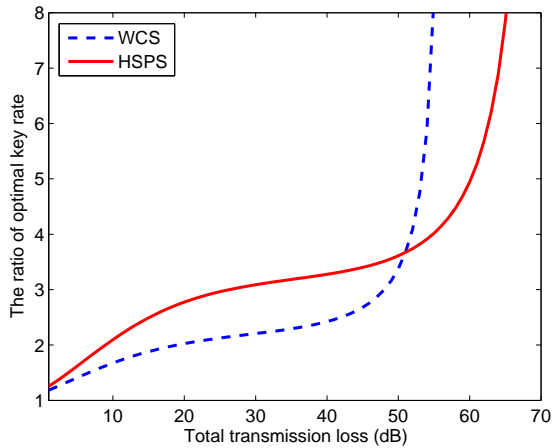}\\
  \caption{(Color online) The ratio of the optimal key rates between the estimations obtained by using 4-intensity and 3-intensity decoy state methods versus the total channel transmission loss for MDI-QKD with WCS and HSPS. We set $\mu_1=\nu_1=0.1$ decoy states, and $\mu_2=\nu_2$.}\label{MDIROpt4Over3}
\end{figure}

\begin{figure}
  \includegraphics[width=240pt]{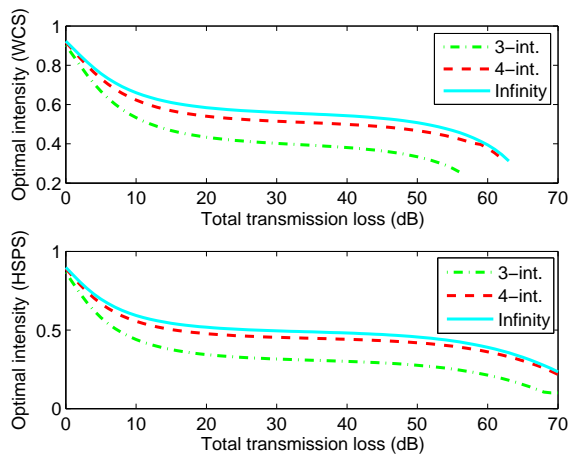}\\
  \caption{(Color online) The optimal intensity of the signal states versus the total channel transmission loss using 3-intensity and 4-intensity decoy state methods for MDI-QKD with WCS and HSPS. We set $\mu_1=\nu_1=0.1$ for  decoy states, and $\mu_2=\nu_2$.}\label{MDIOptmu}
\end{figure}

\begin{table}%[!hbp]
\caption{\label{tabPara}List of experimental parameters used in numerical simulations: $e_0$ is the error rate of background, $e_d$ is the misalignment-error probability; $p_d$ is the dark count rate of UTP's per detector; $f$ is the error correction inefficiency; $\eta_v$ is the detection efficiency of Alice and Bob's detector; $p_{dv}$ is the dark count rate of Alice and Bob's detector.}
\begin{ruledtabular}
\begin{tabular}{cccccc}
  $e_0$ & $e_d$ & $p_d$ & $f$ & $\eta_v$ & $p_{dv}$  \\
  \hline
  0.5 & 1.5\% & $3.0\times 10^{-6}$ & 1.16 & 0.75 & $1.0\times 10^{-6}$ \\
\end{tabular}
\end{ruledtabular}
\end{table}

In this section, we will present some numerical simulations to comparing our results with the results obtained by using 3-intensity decoy state method for MDI-QKD~\cite{Wang3int}. As discussed before, we know that the methods presented in this paper does not only apply to the weak coherent sources (WCS). Actually, it can be used to estimate the final key rate for any sources that satisfy the condition given by Eqs.(\ref{MDIcondxy}-\ref{MDIcondxyz}). Below for simplicity, we consider the following two cases. In the first case, we suppose that Alice and Bob use the WCS. In the second one, we suppose they use the heralded single-photon sources (HSPS) with possion distributions~\cite{wangArxiv}. The UTP locates in the middle of Alice and Bob, and the UTP's detectors are identical, i.e., they have the same dark count rate and detection efficiency, and their detection efficiency does not depend on the incoming signals. We shall estimate what values would be probably observed for the gains and error rates in the normal cases by the linear models as in~\cite{wang05,ind2,WangModel}:
\begin{eqnarray*}
  |n\rangle\langle n| = \sum_{k=0}^n C_n^k \xi^k (1-\xi)^{n-k}|k\rangle\langle k|
\end{eqnarray*}
where $\xi^k$ is the transmittance for a distance from Alice to the UTP.  For fair comparison, we use the same parameter values used in~\cite{ind2} for our numerical evaluation, which follow the experiment reported in~\cite{UrsinNP2007}. For simplicity, we shall put the detection efficiency to the overall transmittance $\eta=\xi^2 \zeta$. We assume all detectors of UTP have the same detection efficiency $\zeta$ and dark count rate $p_d$. In the second case with HPSP, we assume all detectors of Alice and Bob have the same detection efficiency $\eta_v$ and dark count rate $p_{dv}$. The values of these parameters are presented in Table~\ref{tabPara}. With this, by taking the photon-number-cutoff approximation up to 6 photon-number state, the total gains $S_{\mu_i,\nu_j}^{\omega},(\omega=X,Z)$ and error rates $S_{\mu_i,\nu_j}^{\omega}E_{\mu_i,\nu_j}^{\omega},(\omega=X,Z)$ of Alice's intensity $\mu_i$ ($i=0,1,2$ for 3-intensity method, $i=0,1,2,3$ for 4-intensity method) and Bob's intensity $\nu_j$ ($j=0,1,2$ for 3-intensity method, $j=0,1,2,3$ for 4-intensity method) can be calculated. By using these values, we can estimate the lower bounds of yield $s_{11}^{Z}$ with Eq.(\ref{s11L3int}) and Eq.(\ref{s11L4ints}) for 3-intensity and 4-intensity decoy state methods respectively. Also, we can estimate the upper bounds of error rate $e_{11}^{X}$ with Eq.(\ref{e11U3int}) and Eq.(\ref{e11U4int}) for these two decoy state methods respectively. Furthermore, with these parameters, we can estimate the final key rate $R$ of this protocol with Eq.(\ref{KeyRate}). If we fix the densities of the decoy-state sources used by Alice and Bob, the final key rate will change with they taking different intensities for their signal-state pulses. Here, in order to make a rational and effective comparison, we set the intensities of the decoy source in 3-intensity method and the first decoy source in 4-intensity method are the same and $\mu_1=\nu_1=0.1$; let the intensity of the second decoy source in 4-intensity method to be the optimal intensity of signal source in 3-intensity method and assume $\mu_2=\nu_2> \mu_1$~\cite{Note}.

With these preparations, we can conclude that the lower bounds of $s_{11}$ estimated by using the 3-intensity and 4-intensity decoy state methods are the same, i.e., $\underline{s}_{11}=\underline{s}_{11}^{\prime}$. In Fig.\ref{MDIre11U}, we plot the upper bound of $e_{11}$ with different methods. The optimal key rates with different methods for WCS and HSPS are shown in the up and down subfigures respectively in Fig.\ref{MDIROpt}. To see more clearly, in Fig.\ref{MDIrROpt}, we plot the relative value between the optimal key rate obtained with different methods and the asymptotic limit of the infinite decoy-state method. In order to clarify the superiority of the 4-intensity decoy state method, we plot the ratio of the optimal key rate between the results obtained by using 4-intensity and 3-intensity decoy state methods in Fig.\ref{MDIROpt4Over3}. These figures clearly show that our results are better than the pre-existed results. The optimal densities with the optimal key rate versus the total channel transmission loss is given in Fig.\ref{MDIOptmu}. In Fig.\ref{MDIre11U} and Fig.\ref{MDIrROpt}, the black dotted line and green dash-dot line are the results obtained by using 3-intensity decoy state method with WCS and HSPS respectively, the blue dashed line and the red solid line are the results obtained by using 4-intensity decoy state with WCS and HSPS respectively, the thick cyan line are the results obtained by using infinite decoy state method. In Fig.\ref{MDIROpt} and Fig.\ref{MDIOptmu}, the green dotted, the red dashed and the cyan solid lines are the results obtained by using 3-intensity, 4-intensity and the infinite decoy state methods respectively. In Fig.\ref{MDIROpt4Over3}, the blue dashed lines are obtained with WCS, the red solid lines are obtained with HSPS.

\section{Concluding Remark}
In conclusion, we show how to tightly formulate the upper bound of the phase-flip errors in decoy state method for the regular BB84 protocol and MDI-QKD. Our result compressed the bound to about a quarter of known result for MDI-QKD with WCS, and even about one fifth with HSPS. To achieve the result, we only need 4-intensity decoy state method. These methods can be applied to the recently proposed protocols with imperfect single-photon source such as the coherent states or the heralded states from the parametric down conversion. Based on this, we find that the key rate is improved by more than 100\% with WCS, and even more than 200\% with HSPS.

{\bf Acknowledgement:}
We acknowledge
the support from the 10000-Plan of Shandong province,
the National High-Tech Program of China Grants
No. 2011AA010800 and No. 2011AA010803 and NSFC
Grants No. 11174177 and No. 60725416.

\begin{center}
  {\bf Appendix A. Eqs.(\ref{BB84Condxy},\ref{BB84Condyz},\ref{BB84Condxyz},\ref{MDIcondxy},\ref{MDIcondxyz}) with the imperfect sources used in practice}
\end{center}

We know that the state emits from a parametric down-conversion (PDC) source is [17,18]
\begin{equation*}
  \rho_{l}=\sum_{k}a_k^l \oprod{k}{k},
\end{equation*}
with $a_k^l=e^{-l}{l^k}/{k!}$ or $a_k^l={l^k}/{(l+1)^{k+1}}$ where $\ket{k}$ represents an $k$-photon state, $l$ is the intensity (average photon number) of $\rho_l$. Firstly, in this appendix, we will prove that the assumptions given by Eqs.(\ref{BB84Condxy},\ref{BB84Condyz},\ref{BB84Condxyz}) are satisfied by the PDC source. In the 4-intensity protocol, Alice has 3 different no-vacuum sources which are denoted by $\rho_{x},\rho_y,\rho_z$ with $0<x<y<z$.

In the case with $a_k^l=e^{-l}{l^k}/{k!}$, we have
\begin{equation*}
  \frac{a_k^y}{a_k^x}=e^{x-y}\frac{y^k}{x^k}, \quad \frac{a_k^z}{a_k^y}=e^{y-z}\frac{z^k}{y^k}, \quad (k\geq 0).
\end{equation*}
Then we can easily prove the conclusions in Eqs.(\ref{BB84Condxy},\ref{BB84Condyz}) with $x<y<z$. In order to prove the result presented in Eq.(\ref{BB84Condxyz}), we need the following lemma.
\\
\noindent\textbf{Lemma 1.}
{\em{For any two natural number $m,n$ with $m>n\geq 1$, $\mathcal{F}(v)=\frac{1-v^m}{1-v^n}$ is a monotone increasing function in the domain $v\in(0,1)$.}}

The function $\mathcal{F}(v)=\frac{1-v^m}{1-v^n}$ can be rewritten into
\begin{eqnarray*}
  \mathcal{F}(v)&=&\frac{\sum_{k=0}^{m-1}v^k}{\sum_{k=0}^{n-1}v^k}\\
  &=&1+v^n\frac{\sum_{k=0}^{m-n-1}v^k}{\sum_{k=0}^{n-1}v^k} =1+\frac{\sum_{k=0}^{m-n-1}v^k}{\sum_{k=1}^{n}1/v^k}.
\end{eqnarray*}
This predicts that the function $\mathcal{F}(v)$ is monotone increasing with $m>n\geq 1$ in the domain $v\in(0,1)$.

With the definition of $\mathcal{G}(i,j,k)$ in Eq.(\ref{Gdef}), we have
\begin{eqnarray*}
  & &e^{x+y-2z}\mathcal{G}(i,j,k)\\
  &=&[x_r^i-x_r^j][y_r^j-y_r^k]-[y_r^i-y_r^j][x_r^j-x_r^k] \\
  &=&x_r^i y_r^j[1-x_r^{j-i}][1-y_r^{k-j}] -y_r^i x_r^j[1-y_r^{j-i}][1-x_r^{k-j}],
\end{eqnarray*}
with $x_r=x/z$ and $y_r=y/z$.
If $x<y<z$, and $k-j\geq j-i\geq 0$, we get
\begin{equation*}
  \frac{x_r^i y_r^j}{y_r^i x_r^j}=\frac{x^i}{z^i}\frac{y^j}{z^j}\cdot \frac{z^i}{y^i}\frac{z^j}{x^j}=\left(\frac{y}{x}\right)^{j-i}\geq 1,
\end{equation*}
and
\begin{eqnarray*}
  & &(1-x_r^{j-i})(1-y_r^{k-j}) -(1-y_r^{j-i})(1-x_r^{k-j}) \\
  &=&(1-x_r^{j-i})(1-y_r^{j-i})\left(\frac{1-y_r^{k-j}}{1-y_r^{j-i}}-\frac{1-x_r^{k-j}}{1-x_r^{j-i}}\right)\geq 0.
\end{eqnarray*}
In the last step, we have used Lemma 1. With these relations, we can finish the proof of Eq.(\ref{BB84Condxyz}).

In the case with $a_k^l=l^k/(l+1)^{k+1}$, we have
\begin{equation*}
  \frac{a_k^y}{a_k^x}=\frac{x+1}{y+1}\left(\frac{xy+y}{xy+x}\right)^k, \quad \frac{a_k^z}{a_k^y}=\frac{y+1}{z+1}\left(\frac{yz+z}{yz+y}\right)^k,
\end{equation*}
for all $k\geq 1$.
Then we can easily prove the conclusions in Eqs.(\ref{BB84Condxy},\ref{BB84Condyz}) with $x<y<z$. By introducing
\begin{equation*}
  \tilde{x}_r=\frac{xz+x}{xz+z}, \quad \tilde{y}_r=\frac{yz+y}{yz+z},
\end{equation*}
we find out
\begin{eqnarray*}
  & &\frac{(1+x)(1+y)}{(1+z)^2}\mathcal{G}(i,j,k)\\
  &=&[\tilde{x}_r^i-\tilde{x}_r^j][\tilde{y}_r^j-\tilde{y}_r^k] -[\tilde{y}_r^i-\tilde{y}_r^j][\tilde{x}_r^j-\tilde{x}_r^k] \\
  &=&\tilde{x}_r^i \tilde{y}_r^j[1-\tilde{x}_r^{j-i}][1-\tilde{y}_r^{k-j}] -\tilde{y}_r^i \tilde{x}_r^j[1-\tilde{y}_r^{j-i}][1-\tilde{x}_r^{k-j}].
\end{eqnarray*}
If $x<y<z$, with Lemma 1, we can prove that $\mathcal{G}(i,j,k)\geq 0$ when $k-j\geq j-i\geq 0$. This complete the proof of Eq.(\ref{BB84Condxyz}).

Similarly, we can prove that those assumptions in Eqs.(\ref{BB84Condxy},\ref{BB84Condyz},\ref{BB84Condxyz}) can be fulfilled by the heralded single-photon sources (HSPS) with possion or thermal distributions~\cite{wangArxiv}. When we consider the 4-intensity decoy state method for MDI-QKD, the assumptions presented in Eqs.(\ref{MDIcondxy}-\ref{MDIcondxyz}) can be fulfilled if Alice and Bob choose PDC sources or HSPS.

\begin{center}
  {\bf Appendix B. The derivation of the simplified forms of $s_1^{\prime}$ and $s_{11}^{\prime}$}
\end{center}

As discussed in section~\ref{SecBB84}, the lower bound of $s_1$ can be estimated by Eq.(\ref{BB84s1L3}) when Alice use three different sources $\rho_o,\rho_x$ and $\rho_y$. Furthermore, in this case, we can write $s_1$ into the following form with $\underline{s}_1(x,y)$
\begin{equation*}
  s_1=\underline{s}_1(x,y)+\sum_{m\geq 3}f_{s_1}^{(x,y)}(m)s_m,
\end{equation*}
where
\begin{equation*}
  f_{s_1}^{(x,y)}(m)=\frac{a_2^x a_m^y-a_2^y a_m^x}{a_1^x a_2^y -a_1^y a_2^x}, \quad (m\geq 3).
\end{equation*}
Similarly, if Alice choose sources $\rho_o,\rho_x,\rho_z$ and $\rho_o,\rho_y,\rho_z$, then $s_1$ can also be expressed into
\begin{eqnarray*}
  s_1&=&\underline{s}_1(x,z)+\sum_{m\geq 3}f_{s_1}^{(x,z)}(m)s_m, \\
  s_1&=&\underline{s}_1(y,z)+\sum_{m\geq 3}f_{s_1}^{(y,z)}(m)s_m,
\end{eqnarray*}
with
\begin{equation*}
  f_{s_1}^{(x,z)}(m)=\frac{a_2^x a_m^z-a_2^z a_m^x}{a_1^x a_2^z -a_1^z a_2^x}, \quad f_{s_1}^{(y,z)}(m)=\frac{a_2^y a_m^z-a_2^z a_m^y}{a_1^y a_2^z -a_1^z a_2^y},
\end{equation*}
respectively. By calculation, we have
\begin{eqnarray*}
  f_{s_1}^{(x,y)}(m)-f_{s_1}^{(x,z)}(m)&=&-\frac{a_2^x a_1^z a_2^z a_3^z \mathcal{G}(1,2,m)} {(a_1^x a_2^y-a_1^y a_2^x)(a_1^x a_2^z-a_1^z a_2^x)}, \\
  f_{s_1}^{(x,z)}(m)-f_{s_1}^{(y,z)}(m)&=&-\frac{a_2^z a_1^z a_2^z a_3^z \mathcal{G}(1,2,m)} {(a_1^x a_2^z-a_1^z a_2^x)(a_1^y a_2^z-a_1^z a_2^y)}.
\end{eqnarray*}
According to the conditions given by Eqs.(\ref{BB84Condxy},\ref{BB84Condyz},\ref{BB84Condxyz}), we can easily prove that \begin{equation*}
  f_{s_1}^{(x,y)}(m)\leq f_{s_1}^{(x,z)}(m)\leq f_{s_1}^{(y,z)}(m),
\end{equation*}
for all $m\geq 3$. So we have
\begin{equation*}
  \underline{s}_1(x,y)\geq \underline{s}_1(x,z) \geq \underline{s}_1(y,z).
\end{equation*}
This completes the proof of Eq.(\ref{BB84s1L4s}).

Now we commit to prove Eq.(\ref{s11L4ints}) for MDI-QKD. By choosing any two different no-vacuum sources $\rho_{l_{A(B)}},\rho_{r_{A(B)}}$ form $\rho_{x_{A(B)}},\rho_{y_{A(B)}},\rho_{z_{A(B)}}$, we have
\begin{equation*}
  s_{11}=\underline{s}_{11}(l,r)+\sum_{(m,n)\in J_1} f_{s_{11}}^{(l,r)}(m,n)s_{mn},
\end{equation*}
with $\underline{s}_{11}(l,r)$ being defined in Eq.(\ref{s11L3int}) by replacing $x,y$ with $l,r$ respectively, and
\begin{widetext}
\begin{equation*}
  f_{s_{11}}^{(l,r)}(m,n)=\frac{a_2^l b_n^l(a_1^l a_m^r-a_1^r a_m^l)(b_1^l b_2^r-b_1^r b_2^l)+ a_m^l b_1^l (a_1^l a_2^r -a_1^r a_2^l)(b_2^l b_n^r -b_2^r b_n^l)}{a_1^l b_1^l (a_1^l a_2^r-a_1^r a_2^l)(b_1^l b_2^r-b_1^r b_2^l)}.
\end{equation*}
\end{widetext}
where $(l,r)\in\{(x,y),(x,z),(y,z)\}$, $J_1=\{(m,n)|m,n\geq 1; m+n\geq 4\}$. In the coming, we will compare the relations among $f_{s_{11}}^{(x,y)}(m,n),f_{s_{11}}^{(x,y)}(m,n),f_{s_{11}}^{(x,y)}(m,n)$. Firstly, we have
\begin{equation*}
  f_{s_{11}}^{(x,y)}(1,n)-f_{s_{11}}^{(x,z)}(1,n)=\frac{-b_2^x b_1^z b_2^z b_n^z \mathcal{H}(1,2,n)}{(b_1^x b_2^y-b_1^y b_2^x) (b_1^x b_2^z-b_1^z b_2^x)},
\end{equation*}
\begin{equation*}
  f_{s_{11}}^{(x,z)}(1,n)-f_{s_{11}}^{(y,z)}(1,n)=\frac{-b_1^z b_2^z b_2^z b_n^z \mathcal{H}(1,2,n)} {(b_1^x b_2^y-b_1^y b_2^x) (b_1^x b_2^z-b_1^z b_2^x)},
\end{equation*}
for all $n\geq 3$. Secondly, we obtain
\begin{equation*}
  f_{s_{11}}^{(x,y)}(n,1)-f_{s_{11}}^{(x,z)}(n,1)=\frac{-a_2^x a_1^z a_2^z a_n^z \mathcal{G}(1,2,n)}{(a_1^x a_2^y-a_1^y a_2^x)(a_1^x a_2^z-a_1^z a_2^x)},
\end{equation*}
\begin{equation*}
  f_{s_{11}}^{(x,z)}(n,1)-f_{s_{11}}^{(y,z)}(n,1)=\frac{-a_1^z a_2^z a_2^z a_n^z \mathcal{G}(1,2,n)}{(a_1^x a_2^y-a_1^y a_2^x)(a_1^x a_2^z-a_1^z a_2^x)},
\end{equation*}
for all $n\geq 3$. In the last case, we get
\begin{widetext}
\begin{equation*}
  f_{s_{11}}^{(x,y)}(m,n)-f_{s_{11}}^{(x,z)}(m,n)=-\frac{a_2^x b_n^x a_1^z a_2^z a_m^z\mathcal{G}(1,2,m)}{b_1^x (a_1^x a_2^y-a_1^y a_2^x) (a_1^x a_2^z-a_1^z a_2^x)} -\frac{ b_2^x a_m^x b_1^z b_2^z b_n^z \mathcal{H}(1,2,n)}{a_1^x (b_1^x b_2^y-b_1^y b_2^x) (b_1^x b_2^z-b_1^z b_2^x)},
\end{equation*}
and
\begin{equation*}
  f_{s_{11}}^{(x,z)}(m,n)-f_{s_{11}}^{(y,z)}(m,n)\leq -\frac{a_2^y b_n^y a_1^z a_2^z a_1^z a_m^z\mathcal{G}(1,2,m)}{a_1^y b_1^y (a_1^x a_2^z-a_1^z a_2^x) (a_1^y a_2^z-a_1^z a_2^y)} -\frac{ a_m^y b_1^z b_2^z b_2^z b_n^z \mathcal{H}(1,2,n)}{a_1^y (b_1^x b_2^z-b_1^z b_2^x) (b_1^y b_2^z-b_1^z b_2^y)},
\end{equation*}
\end{widetext}
for all $m,n\geq 2$. In the lase inequality, we have used the assumption presented in Eq.(\ref{MDIcondxy}). With these relations, we can conclude that
\begin{equation*}
  f_{s_{11}}^{(x,y)}(m,n)\leq f_{s_{11}}^{(x,z)}(m,n)\leq f_{s_{11}}^{(y,z)}(m,n),
\end{equation*}
for any $(m,n)\in J_1$ under the conditions in Eqs.(\ref{MDIcondxy}-\ref{MDIcondxyz}). This completes the proof of Eq.(\ref{s11L4ints}).

%\clearpage

\end{document}